\documentclass[12pt]{article} 
\usepackage{esfconf}
\usepackage{latexsym} 
\def\a{\alpha} 
\def\b{\beta} 
\def\g{\gamma}  
\def\d{\delta}

\def\m{\mu}  
\def\n{\nu}  
\def\r{\rho}

\def\cA{{\cal A}}
 
\def\pa{\partial}
 
\def\ll{\left} 
\def\rr{\right} 
\def\be{\begin{equation}}  
\def\ee{\end{equation}}  
\def\beq{\begin{eqnarray}}  
\def\eeq{\end{eqnarray}}  
\def\nn{\nonumber}

\newtheorem{theorem}{Theorem}
\begin{document} 

\begin{flushright}

IHES/P/00/62

ULB-TH-00/16

HEP-TH/0009109

\end{flushright}

\vspace{1.2cm}

\title{No consistent 
cross-interactions for a collection of massless spin-2 fields}

\authors{Nicolas Boulanger\adref{1},  
Thibault Damour\adref{2}, Leonardo Gualtieri\adref{1} and 
Marc Henneaux\adref{1,3}}  

\addresses{\1ad Physique Th\'eorique et Math\'ematique,  Universit\'e Libre 
de Bruxelles,  C.P. 231, B-1050  Bruxelles, Belgium  \\
\2ad
Institut des Hautes Etudes Scientifiques,  35, route de 
Chartres,  F-91440 Bures-sur-Yvette, France  \\ 
\3ad Centro de Estudios Cient\'{\i}ficos, Casilla 1469, Valdivia, Chile }

\maketitle
 
\begin{abstract} 
We report  a no-go theorem excluding  
consistent cross-couplings for a collection 
of massless, spin-2 fields described, in the free limit, by the sum 
of Pauli-Fierz actions (one for each field). 
We show that, in spacetime dimensions
$>2$,  there is no consistent coupling, with at most 
two derivatives of the fields, that can mix the various 
``gravitons''.  The only possible deformations are given by the 
sum of individual Einstein-Hilbert actions (one for each field) 
with cosmological 
terms.  
Our approach is based on the BRST-based deformation point of 
view$^\dagger$.
\end{abstract}

\vspace{5cm}

\noindent
$^\dagger$ {\footnotesize To appear in the 
Proceedings of the Meetings ``Spring School
in QFT and Hamiltonian Systems" (Calimanesti, Romania, 2-7 May 2000)
and ``Quantization, Gauge Theory and 
Strings" (Moscow, Russia, 5-10 June 2000).} 
\newpage 
 
\section{Introduction} 

One striking feature of the Einstein theory of gravity is 
that it involves a single massless spin-two field.  We report
here 
results obtained recently \cite{Boulangeretal}
showing that, in fact, this is not an accident : theories involving different
types of gravitons with non trivial, consistent, cross-interactions simply
do not exist.  This no-go theorem holds under the assumption that (i) the
Lagrangian contains no more than two derivatives of the massless spin-2 fields 
$\{h_{\mu \nu}^a\}$ ($a = 1, \cdots, N$); (ii) the interactions can be
continuously switched on; and (iii) in the limit of no interaction, the
action reduces to the sum of 
one  Pauli-Fierz action \cite{Fierz:1939ix}   
for each field $h_{\mu \nu}^a$, i.e. 
\beq 
\label{startingpoint} 
S_0[h_{\mu \nu}^a] &=& \sum_{a = 1}^N \int d^n x \ll[ 
-\frac{1}{2}\ll(\pa_{\m}{h^a}_{\n\r}\rr)\ll(\pa^{\m}{h^a}^{\n\r}\rr) 
+\ll(\pa_{\m}{h^a}^{\m}_{~\n}\rr)\ll(\pa_{\r}{h^a}^{\r\n}\rr)\rr.\nn\\ 
&&\ll.-\ll(\pa_{\n}{h^a}^{\m}_{~\m}\rr)\ll(\pa_{\r}{h^a}^{\r\n}\rr) 
+\frac{1}{2}\ll(\pa_{\m}{h^a}^{\n}_{~\n}\rr) 
\ll(\pa^{\m}{h^a}^{\r}_{~\r}\rr)\rr] 
\eeq 
(spacetime
indices are raised and lowered with the flat Minkowskian metric
$\eta_{\mu \nu}$, for which we use a ``mostly plus" signature).

The free action (\ref{startingpoint}) 
is invariant under the linear gauge transformations, 
$\delta_\epsilon h^a_{\mu \nu} = \partial_\mu \epsilon_\nu^a 
+ \partial_\nu \epsilon_\mu^a $
where the $\epsilon_\nu^a$ are arbitrary functions.  These transformations 
are abelian and irreducible.  The Pauli-Fierz action is in fact the linearized 
Einstein action and describes a pure spin-2 system (no spin 1 or 0 included). 
 
The equations of motion are 
$H_a^{\mu \nu} = 0 $,
where $H^a_{\mu \nu}$ is the linearized Einstein tensor.
The Noether identities expressing the gauge invariance of the free 
action are
$\partial_\nu H^{a \mu \nu} = 0 $
(linearized Bianchi identities).  The gauge symmetry removes unwanted 
unphysical states. 
 
The problem of introducing consistent interactions for a collection 
of massless spin-2 fields  
is that of adding local interaction terms to the 
action (\ref{startingpoint}) 
while modifying at the same time  
the original gauge symmetries if necessary, in such a 
way that the modified action is invariant under the modified gauge 
symmetries. Since we are interested in the 
classical theory, we shall also demand that the interactions contain no 
more than two derivatives so that the nature of the differential 
equations for $h^a_{\mu \nu}$ is unchanged.
We shall, however, make no assumption on the 
polynomial order of the fields in the Lagrangian or in the gauge 
symmetries. 
 
In an  interesting work \cite{Wald1},  Cutler and Wald have 
proposed multi-graviton theories with cross-interactions  
based on associative, commutative algebras.  These authors 
arrived at these structures by focusing 
on the modified gauge transformations and their algebra, but 
did not analyze the extra conditions that must be imposed 
on the modified gauge symmetries if these are to be compatible  
with a Lagrangian having the free field limit prescribed above. 
 
Some explicit examples of 
Lagrangians that realize the Cutler-Wald algebraic 
structures have been constructed 
in \cite{Wald2} and \cite{ovrut}, but none of these has 
the correct free field limit.  In fact, their free field 
limit does involve a sum of 
Pauli-Fierz Lagrangians, but some of the 
``gravitons'' come with the wrong sign and thus, the energy 
of the theory is unbounded from below.  
To our knowledge, 
the question of whether other examples of (real) Lagrangians realizing 
the Cutler-Wald structure could exist and whether some of them 
would have the correct free field limit was left open. 
  
Motivated by these developments, we have re-analyzed the 
question of consistent interactions for a collection 
of massless spin-2 fields by imposing from the outset that the 
deformed Lagrangian should have the free field limit 
(\ref{startingpoint}).  It turns out that this requirement forces 
one additional condition on the Cutler-Wald algebra defining 
the interaction, namely, that it  
should be symmetric in the scalar product defined by the 
free Lagrangian.  This extra constraint is quite stringent and 
implies that the algebra is the direct sum of one-dimensional ideals. 
This eliminates all the cross-interactions.
The only consistent 
deformation (within the context of no more than two derivatives) 
that the free theory based on (\ref{startingpoint}) admits is the  
sum of one Einstein-Hilbert action (with a possible cosmological 
term) for each spin-two field, 
\be 
S[g^a_{\mu \nu}] = \sum_a \frac{2}{\kappa_a^2}\int d^nx 
(R^a-2 \Lambda^a) \sqrt{-g^a}, \; g^a_{\mu \nu} = 
\eta_{\mu \nu} + \kappa^a h^a_{\mu \nu} 
\label{sum} 
\ee 
where $R^a$ is the scalar curvature of $g^a_{\mu \nu}$ and 
$g^a$ its determinant.  There is no other possibility.
[Some sectors may remain undeformed; for $\kappa_a = 0$, the
action reduces to the Pauli-Fierz term plus a possible cosmological
term $\lambda^a h^{a \mu}_{\; \; \mu}$.] 
  
We present here the main ideas underlying our no-go
theorem. Details and
proofs may be found in \cite{Boulangeretal}.

\section{Cohomological reformulation} 

Our approach is based on the BRST reformulation of the problem \cite{bh}, in
which consistent couplings define deformations of the
solution of the so-called ``master equation''.
The advantage of this approach
is that it clearly organizes the calculation of the non-trivial consistent
couplings in terms of cohomologies which are known or easily computed.
These cohomologies are in fact interesting in themselves, besides
their occurence in the consistent interaction problem.
The use of BRST techniques
somewhat streamlines the derivation, which would otherwise
be more cumbersome.        

Let us thus write down first 
the solution of the master equation for a collection of free, 
spin-2, massless fields.
According to the general rules \cite{bv1}, the spectrum of
fields, ghosts 
and antifields is given by :
(i) the fields $h^a_{\a\b}$, with ghost number and antighost number zero; 
(ii) the ghosts $C^a_{\a}$, with ghost number one and antighost number zero; 
(iii) the antifields $h^{* \a\b}_{a}$, with ghost  
number minus one and antighost  
number one; 
and (iv) the antighosts $C^{* \a}_{a}$, with ghost number minus  
two and antighost  
number two. 

While the ghost number assignments are rather standard, the introduction
of another grading, namely, the antighost number, may appear to be a
bit artificial.  It turns out, however, that this is not so.  The
antighost number (also called antifield number) is not only technically
useful, but it also enables one to keep track of terms with different
meanings in the master equation.  We shall come back to this
point at the end of this section.
 
According to the prescriptions of \cite{bv1},
the solution of the master equation for the free theory is, 
$W_0 = S_0 + \int d^nx \, h^{* \a\b}_{a} (\partial_\a C^a_\b 
+ \partial_\b C^a_\a)$, 
from which we get the BRST differential $s$ of the free 
theory as  
$s \cdot = (W_0, \cdot)$.  Here, $(,)$ is the antibracket.
Explicit calculations show that $s$ splits as $s = \delta + \gamma $
where the action of $\gamma$ and $\delta$ on the variables is  
zero except 
\be 
\g h^a_{\a\b}=2\pa_{(\a}C^a_{\b)}, \;
\d h_a^{*\a\b}=\frac{\d S_0}{\d h^a_{\a\b}}, \;
\d C_a^{*\a}=-2\pa_{\b}h_a^{*\b\a}. \label{defd2} 
\ee 
The decomposition of $s$ into $\delta$ plus $\gamma$ is 
dictated by the antighost number: $\delta$ decreases the 
antighost number by one unit, while $\gamma$ leaves it unchanged. 
One has
$\delta^2 = 0, \; \delta \gamma + \gamma \delta = 0, \; 
\gamma^2 = 0$. 

If one expands the solution $W$ of the master equation $(W,W)=0$ 
for the searched-for interacting theory in powers of the deformation
parameter $g$ (coupling constant), 
$W= W_0 + g W_1 + g^2
W_2 + \cdots$, one finds the conditions $sW_1 \equiv (W_0, W_1)
= 0$ and $(W_1, W_1) = -2 s W_2$ at orders one
and two, respectively.
The first condition expresses that the first-order deformation $W_1$
should be a BRST-cocycle.  Trivial cocycles (of the form
$s K$) define actually
``fake" interactions, in the sense that they can be
absorbed through fields and ghosts redefinitions \cite{bh}.
The second condition expresses 
that $(W_1,W_1)$ - which is easily verified to be
BRST-closed - should be BRST-exact in order for $W_2$ to
exist. 
Since we deal with local functionals, the relevant cohomology groups
are, in terms of the integrands, $H(s\vert d)$ \cite{locality,BBH1}.
Thus, first-order deformations are characterized by elements of
$H^{0,n}(s \vert d)$ (BRST cohomology at ghost number zero and form
degree $n$ for the $n$-form integrand $a$ of the
ghost number functional $W_0 = \int a$); and obstructions to
continuing a given first-order consistent
deformation to order $g^2$ are measured
by $H^{1,n}(s \vert d)$.

In the sequel, we shall compute explicitly $H^{0,n}(s \vert d)$ for
a collection of free, massless spin-2 fields, i.e., we shall determine
all possible first-order consistent interactions.  We shall then determine
the conditions that these must fulfill in order to
be unobstructed at order $g^2$.  These
conditions turn out to be extremely strong and prevent
cross interactions between the various types of gravitons.      

We finally close this section by observing that $W_0$ and $W$
have ghost number zero, but break into various components with different
antighost numbers.  For instance, $W_0$ has a piece with antighost number
zero and another with antighost number one. The first piece is the
classical action, while the second contains the information about the
gauge symmetries.  This feature is quite general: the antighost number
zero component of the solution of the master equation is the classical
action, the antighost number one component contains the information
about the gauge symmetries while the antighost number two component
contains the information about the gauge algebra.  The absence of such a term
in $W_0$ reflects the fact that the gauge algebra of the
free theory is abelian.  By deforming the solution of the master equation,
one deforms everything (action, gauge transformations, gauge algebra)
at once; but one can recover the detailed information by splitting $W$ 
according to the antighost number.

\section{Cohomology of $\gamma$} 
\label{cohoofg} 
 
To compute the consistent, first order deformations, i.e., 
$H(s \vert d)$, we need $H(\gamma)$ and $H(\delta \vert d)$. 
We start with $H(\gamma)$, which is rather easy. 
 
As it is clear from its definition, $\gamma$ is related to the 
gauge transformations.  Acting on anything, it gives zero, except 
when it acts on the spin-2 fields, on which it gives a gauge transformation 
with gauge parameters replaced by the ghosts. 
 
The only gauge-invariant objects that one can construct out of 
the gauge fields $h^a_{\mu \nu}$ and their derivatives are the 
linearized curvatures $K^a_{\a \b \m \n}$ and their derivatives. 
The antifields and their derivatives are also $\gamma$-closed.  
The ghosts and their derivatives are $\gamma$-closed as well but 
their symmetrized first order derivatives are $\gamma$-exact , as
are all their subsequent derivatives since 
$\pa_{\a\b}C^a_{\g}=\frac{1}{2}\,\g\ll(\pa_{\a}h^a_{\b\g}+\pa_{\b}h^a_{\a\g} 
-\pa_{\g}h^a_{\a\b}\rr)$. 
 
It follows straightforwardly from these observations 
that the $\g$-cohomology is generated 
by the linearized curvatures, the antifields and all their derivatives, as 
well as by the 
ghosts $C^a_\mu$ and their antisymmetrized first-order derivatives 
$\pa_{[\m}C^a_{\n]}$.  More precisely, let $\{\omega^I\}$ be a basis 
of the space of polynomials in the $C^a_\mu$ and $\pa_{[\m}C^a_{\n]}$ (since 
these variables anticommute, this space is finite-dimensional).  One has: 
\be 
\label{alphaomega} 
\g a = 0 \Rightarrow a =  
\a_J\ll([K],[h^*],[C^*]\rr)\omega^J\ll(C^a_{\m},\pa_{[\m}C^a_{\n]}\rr) 
+ \gamma b \,, 
\ee 
where the notation $f([m])$ means that the function $f$ depends 
on the variable $m$ and its subsequent derivatives up to a finite order. 
If $a$ has a fixed, finite ghost number, then $a$ can 
only contain a finite number of antifields.  If we assume in addition 
that $a$ has a bounded number of derivatives, as we shall do from now on, 
then, the $\a_J$ are polynomials.
 
In the sequel, the polynomials $\a_J\ll([K],[h^*],[C^*]\rr)$ in the 
linearized curvature $K^a_{\a \b \m \n}$, the antifields $h^{* \m \n}_a$ 
and $C^{* \m}_a$, as well as all their derivatives, will be 
called ``invariant polynomials''.  They may of course  
have an extra, unwritten, 
dependence on $dx^\m$, i.e., be exterior forms. 
At zero antighost number, the invariant polynomials are 
the polynomials in the linearized curvature $K^a_{\a \b \m \n}$ 
and its derivatives. 
 
We shall need the following theorem on the cohomology of $d$ in the 
space of invariant polynomials. 
\begin{theorem}\label{2.2} 
In form degree less than $n$ and in 
antighost number strictly greater than $0$, 
the cohomology of $d$  
is trivial in the space of invariant 
polynomials. 
\end{theorem} 
That is to say, if $\a$ is an invariant polynomial
with $antigh(\a) > 0$, the equation 
$d \a = 0$  implies   
$ \a = d \b$ where $\b$ is also an invariant polynomial. 
For the proof, see \cite{Boulangeretal}.
 
\section{Cohomology of $\delta$ 
modulo $d$} 
\label{characteristic} 

The next cohomology that we shall need is $H(\delta \vert d)$ in the space 
of local forms that do not involve the ghosts ($H(\delta \vert d)$ 
is trivial in the space of forms with positive ghost number 
\cite{locality}). 
This cohomology has an interesting interpretation in terms of conservation  
laws (\cite{BBH1} 
for more information).
 
The following vanishing theorems can be proven: 
\begin{theorem} 
\label{vanishing} 
The cohomology groups $H^n_p(\delta \vert d)$ vanish 
in antighost number strictly greater than $2$, 
\be 
H^n_p(\delta \vert d) = 0 \, \hbox{ for } p>2. 
\ee 
\end{theorem} 
The proof of this theorem 
is given in \cite{BBH1} and follows from the fact that 
linearized gravity is a linear, irreducible, gauge theory. 

In antighost number two, the cohomology is also completely known, 
\begin{theorem} 
\label{conservation2} 
A complete set of representatives of $H^n_2(\delta \vert d)$  
is given by the antifields $C^{*\m}_a$ conjugate to the 
ghosts, i.e.,  
\be 
\delta a^n_2 + da^{n-1}_1 = 0 \Rightarrow a^n_2 
= \lambda^a_\m C^{*\m}_a dx^0 dx^1 \cdots dx^{n-1} 
+ \delta b^n_3 + d b^{n-1}_2 
\ee 
where the $\lambda^a_\m$ are constant. 
\end{theorem} 
 
For the proof, see \cite{Boulangeretal}.

We have discussed so far the cohomology of $\delta$ modulo $d$ in the space 
of arbitary functions of the fields $h^a_{\m \n}$, the antifields, and 
their derivatives.  One can also study $H^n_k(\delta \vert d)$ in the 
space of invariant polynomials in these variables, which involve 
$h^a_{\m \n}$ and its derivatives only through the linearized 
Riemann tensor and its derivatives (as well as 
the antifields and their derivatives). The above theorems remain unchanged 
in this space.  This is a consequence of 
\begin{theorem} 
Let $a$ be an invariant polynomial.  Assume that $a$ is $\delta$ trivial modulo 
$d$ in the space of all (invariant and non-invariant) polynomials, 
$a = \delta b + dc$.  Then, $a$ is also $\delta$ trivial modulo 
$d$ in the space of invariant polynomials, i.e., one can assume 
without loss of generality that $b$ and $c$ are invariant polynomials. 
\end{theorem} 
The proof is given in \cite{Boulangeretal}.  
 
\section{Construction of the general gauge theory of interacting 
gravitons by means of cohomological techniques} 
\label{hard} 

To compute $H^{n,0}(s \vert d)$, we shall use an expansion according to the 
antighost number, as in \cite{BBH2}. 
Let $a$ be a solution of  
$sa + db = 0 $
with ghost number zero. 
One can expand $a$ as 
$a = a_0 + a_1 + \cdots a_k $
where $a_i$ has antighost number $i$ (and ghost number zero). 
Without loss of generality, one can assume that this expansion 
stops at some finite value of the antighost number. 
This was shown in \cite{BBH2} (section 3), under the 
sole assumption that the first-order deformation of the 
Lagrangian $a_0$ has a finite (but otherwise 
arbitrary) derivative order. 
 
The previous theorems on the characteristic cohomology imply that 
one can remove all components of $a$ with antighost number 
greater than or equal to 3. Indeed, the (invariant) 
characteristic cohomology in degree $k$ measures precisely the obstruction 
for removing from $a$ the term $a_k$ of antighost number $k$ 
(see \cite{Boulangeretal}).  Since $H^n_k(\delta \vert d)$ vanishes
for $k>2$, one can 
assume 
$a = a_0 + a_1 + a_2$ and
$b = b_0 + b_1$ \cite{Boulangeretal}. 
Inserting these expressions
in $sa + db = 0$, we get  
$\d a_1+\g a_0=db_0$, 
$\d a_2+\g a_1=db_1$ and
$\g a_2=0$.
Let us recall the meaning of the various terms in $a$ : 
$a_0$ is the deformation of the lagrangian; 
$a_1$ captures the information about the deformation 
of the gauge transformations; 
while $a_2$ contains the information about the deformation 
of the gauge algebra. 
 
\subsection{Determination of $a_2$} 
\label{deterofa2} 
\par 
As we have seen in section \ref{cohoofg}, 
the general solution of $\g a_2 = 0$ reads, 
modulo trivial terms, 
$a_2 = \sum_J \alpha_J \omega^J $,
where the $\alpha_J$ are invariant polynomials 
(see (\ref{alphaomega})).  A necessary  
(but not sufficient) condition for  
$a_2$ to be also
a solution of $\d a_2 + \g a_1 + db_1 = 0$, so that $a_1$ exists, 
is that $\alpha_J$ be a non trivial element of 
$H^n_2(\d\vert d)$ \cite{Boulangeretal}. 
Thus, the polynomials $\a_J$ must be linear combinations 
of  the 
antighosts $C^*_{\a a}$.  
The monomials $\omega^J$ have ghost number two; so they can be 
of only three possible types, namely,  
$C^a_{\a}C^b_{\b}$, $C^a_{\a}\pa_{[\b}C^b_{\g]}$
and $\pa_{[\a}C^a_{\b]}\pa_{[\g}C^b_{\d]}$. 
They should be combined with $C^{*a}_{\a}$ to form $a_2$. 
By Poincar\'e (and PT) invariance, the only possibility is to take 
$C^a_{\a}\pa_{[\b}C^b_{\g]}$, which  
yields
$a_2'=- C^{*\b}_aC^{\a b}\pa_{[\a}C^c_{\b]}a^a_{bc}$. 
Notice that the constants $a^a_{bc}$ are introduced here as the  
constants on which the general solution $a_2$  
depends. 
 
The $a^a_{bc}$ can be identified with the 
structure constants of a $N$-dimensional 
algebra ${\cal A}$.  Let $V$ be an ``internal" vector space 
of dimension $N$; we define a product in $V$ through 
\be 
(x \cdot y)^a = a^a_{bc} x^b y^c, \; \; \forall x,y \in V. 
\label{defofproduct} 
\ee 
The vector space $V$ equipped with this product defines the algebra 
${\cal A}$.  At this stage, ${\cal A}$ has no particular further 
structure.  Extra conditions will arise, however, from the demand that 
$a$ (and not just $a_2$) exists and defines a deformation 
that can be continued to all orders.  We shall recover in this manner 
the conditions found in \cite{Wald1}, plus one additional condition 
that will play a crucial role. 
 
We redefine $a_2$ by adding a $\g$--exact term to $a_2'$, in order to make  
the subsequent calculations simpler: 
\beq 
\label{a2} 
a_2=C^{*\b}_{a}C^{\a b}\pa_{\b}C^c_{\a}a^a_{bc}
 = a_2'+\g\ll(\frac{1}{2}C^{*\b}_{a}C^{\a b}h^c_{\a\b}a^a_{bc}\rr)\,. 
\eeq 
In terms of the algebra of the gauge transformations, this term $a_2$ 
implies that the gauge parameter $\zeta^{a \m}$ corresponding to the 
commutator of two gauge 
transformations with parameters 
$\xi^{a \m}$ and $\eta^{a \m}$ is given by 
\be 
\zeta^{a \m} = a^a_{bc} [\xi^b, \eta^c]^\m 
\label{commutatorgaugetransf} 
\ee 
where $[,]$ is the Lie bracket of vector fields. 
It is worth noting that at this stage, we have not used any a priori 
restriction on the number of derivatives (except that it is finite). 
The assumption that the interactions contain at most two derivatives 
will only be needed below.  Thus, the fact that $a$ stops at 
$a_2$, and that $a_2$ is given by (\ref{a2}) 
is quite general. 
 
Differently put: to first-order in the coupling constant, the 
deformation of the algebra of the spin-2 gauge symmetries is 
universal and given by (\ref{a2}).  There is no other possibility. 
 
\subsection{Determination of $a_1$} 
\label{deterofa1} 
\par 
In order to find $a_1$ we have to solve the equation 
$\d a_2+\g a_1=db_1$. 
As shown in \cite{Boulangeretal}, this equation for $a_1$ has a solution
if and only if
\be 
a^a_{bc}=a^a_{(bc)}\,, 
\label{commutativity} 
\ee 
so that the  algebra ${\cal A}$ defined by the 
$a^a_{bc}$'s must be commutative.  
This result is not surprising in view of the form of the commutator 
of two gauge transformations since (\ref{commutatorgaugetransf}) 
ought to  be antisymmetric in $\xi^a$ and $\eta^a$.  When (\ref{commutativity}) 
holds, 
$a_1$  is given by
$a_1= - h_a^{*\b\g}C^{\a b}\ll(\pa_{\g}h^c_{\a\b}+\pa_{\b}h^c_{\a\g} 
-\pa_{\a}h^c_{\g\b}\rr)a^a_{bc}$
up to a solution of the ``homogenous" equation 
$\gamma a_1 + db_1 = 0$. 
 
The solutions of the homogeneous equation do not modify the gauge 
algebra (since they have a vanishing $a_2$), but they do modify 
the gauge transformations.  
However, they involve too many derivatives (see \cite{Boulangeretal})
and so, are excluded by our number-of-derivatrives-assumption.

\subsection{Determination of $a_0$} 
\par 
We now turn to the 
determination of $a_0$, that is, to the  
determination of the deformed lagrangian at first 
order in $g$. 
The equation for $a_0$ is 
$\d a_1+\g a_0=db_0$. 
It is shown in \cite{Boulangeretal} that this equation
for $a_0$ has a solution if and only if
\be 
a_{abc} = a_{(abc)} .
\label{allsym} 
\ee   
An algebra which fulfills 
$a_{abc} = a_{cba}$ 
is called hilbertian, or, in the real case considered here, 
``symmetric". 
When (\ref{allsym}) holds, $a_0$ exists and
is given by a cubic expression whose explicit form may be found  
in \cite{Boulangeretal}.·
We have therefore  proven that a gauge theory of interacting spin two fields, 
with a non trivial gauge algebra, is first-order consistent 
if and only if the algebra ${\cal A}$ defined by $a^a_{bc}$, 
which characterizes $a_2$, 
is commutative and symmetric.

Again, there is some ambiguity in $a_0$ since we can add 
to it any solution of the ``homogeneous" equation 
$\gamma \tilde{a}_0 + d \tilde{b}_0 = 0$ without $a_1$.  If one requires that 
$\tilde{a}_0$ has no more than two derivatives
- as done here -, there is only one 
possibility, namely 
$- 2 \tilde{\Lambda}_a^{(1)} h^{a \m}_{\; \m} $
where the $\tilde{\Lambda}_a^{(1)}$'s are constant.  This term 
fulfills 
$\g \tilde{\Lambda}_a^{(1)} h^{a \m}_{\; \m} = \partial_\mu (2  
\tilde{\Lambda}_a^{(1)} \epsilon^{a \mu}) $
and is of course the (linearized) cosmological term. 
There is no other term \cite{Boulangeretal}.

\subsection{The associativity of the algebra from the absence 
of obstructions at second order} 
\par 
The master equation at order two is 
$(W_1,W_1) = - 2 s W_2 $.
Given $W_1 =\int d^nx\,\ll(a_0+a_1+a_2\rr)$, 
$W_2$ exists if and only if 
$(W_1,W_1)$ is BRST-exact.  This happens if and
only if the $a^{a}_{bc}$ fulfill
\cite{Boulangeretal}
\be 
a^a_{d[b}a^d_{f]c}=0\,, 
\ee 
which is the associative property for the algebra ${\cal A}$ defined by 
the $a^a_{bc}$.  Thus, ${\cal A}$ must be commutative, symmetric and 
associative. 
 
\section{Impossibility of cross-interactions} 
\label{obstr} 

Finite-dimensional algebras that are commutative, symmetric and associative 
have a trivial structure:  they are the direct sum of one-dimensional 
ideals. 
 
To see this, one proceeds as follows. 
The algebra operation allows us to associate to every element of the algebra 
$u\in\cA$ a linear operator  
$A(u)\,:\,\cA\longrightarrow\cA $
defined by 
$A(u)v\equiv u\cdot v$. 
In a basis $(e_1,\dots,e_m)$, one has $v=v^ae_a$ and 
$A(u)^c_{~b}=u^aa^c_{ab}$. 
Because of the associative property, the operators $A(u)$ provide 
a representation of the algebra 
$A(u)A(v)=A(u\cdot v) $
and so, since the algebra is commutative, 
$[A(u),A(v)]=0$. 
 
Now, the free Lagrangian defines a scalar product in 
the algebra, 
$(u,v)=\d_{ab}u^av^b$. 
The symmetry property  
$a_{abc}=a_{(abc)} $
expresses that the operators $A(u)$ are all symmetric  
$(u,A(v)w)=(A(v)u,w)$, 
that is, 
$A(u)=A(u)^T$. 
Then the operators $A(u),~u\in\cA$ are diagonalizable by 
a rotation.  Since they are commuting,   
they are simultaneously diagonalizable.  In a basis 
$\{e_1,\dots,e_m\}$ in which they are all diagonal, 
one has $A(e_a)e_b =\a(a,b)e_b$ for some numbers $\a(a,b)$ and thus 
$e_a \cdot e_b = A(e_a)e_b=\a(a,b)e_b= e_b \cdot e_a= 
A(e_b) e_a =\a(b,a)e_a$. 
So 
$\a(a,b) = 0$ unless $a = b$.  
 
Consequently, the structure constants $a^a_{bc}$ of the algebra 
${\cal A}$ vanish whenever two indices are different.  There is no term in 
$W_1$ coupling the various spin-2 sectors, which are therefore completely 
decoupled.  Only self-interactions are possible.  The first-order 
deformation $W_1$ is in fact the sum of Einstein cubic vertices 
(one for each spin-2 field with $\a(a,a) \not=0$) $+$ (first-order) 
cosmological terms. 
 
Once the absence of cross-interactions is established, it
is easy to show that the full Lagrangian is given by the sum
(\ref{sum}) of Einstein actions, which are known to be
solutions of the
deformation problem.  The discussion may be found in \cite{Boulangeretal}.

\section{Conclusions} 
\label{conclusions} 

In this note, we have reported a no-go result on cross-interactions 
between a collection of massless spin-2 fields.   
Our method relies on the antifield approach and uses cohomological 
techniques.                                          
 
Although we restricted the discussion to interactions with at most
two derivatives, the same conclusion seems to hold in general,
except for obvious cross-interactions involving
the linearized curvatures, which do not change the
gauge transformations (see \cite{Boulangeretal} for comments on
this).  Also, standard matter is not expected to alter the discussion
(the case of scalar matter is considered explicitly in
\cite{Boulangeretal}).
The interacting theory 
describes thus parallel worlds, and, in any given world, there is
only one massless spin-2 field.  This massless spin-2 field has
the standard graviton couplings with the fields living
in its world (including itself), in agreement with
the single massless spin-2 field studies of
\cite{Guptaetal}.

The fact that that there is effectively only one type of gravitons
is therefore not a choice but a necessity that adds to its great theoretical
appeal.  This feature is one of the arguments used to rule out
$N>8$ extended supergravity theories, since these would involve
gravitons of different types (besides particles of spin greater than
$2$, whose coupling to gravity is known to be problematic).  Our no-go
theorem extends the analysis of \cite{ArDe}, where the coupling of
one massless spin-2 field to gravity described by Riemannian geometry
was shown to be problematic.
 
We close by noting that one key assumption
underlying our negative result is the presence of
only a finite number of gravitons.  This assumption
was crucially used in showing that the structure of
the algebra ${\cal A}$ was trivial.
If this assumption is relaxed, 
cross-interactions become possible \cite{Reuter}.

\section*{Acknowledgements} 
N.B. is ``Chercheur F.R.I.A.".
M.H. thanks the Institut des Hautes Etudes 
Scientifiques for its kind hospitality. 
The work of N. B., L.G. and 
M.H. is partially supported by the ``Actions de 
Recherche Concert{\'e}es" of the ``Direction de la Recherche 
Scientifique - Communaut{\'e} Fran{\c c}aise de Belgique", by 
IISN - Belgium (convention 4.4505.86) and by 
Proyectos FONDECYT 1970151 and 7960001 (Chile). 
M.H. is grateful to the organizers of the meetings ``Spring School in QFT 
and Hamiltonian Systems" (Calimanesti, Romania, 2-7 May 2000) and
``Quantization, Gauge Theory and Strings" (Moscow, Russia, 5-10 June 2000)
where this work
was presented.


\begin{thebibliography}{99} 
\bibitem{Boulangeretal} N. Boulanger, T. Damour, L. Gualtieri
and M. Henneaux, hep-th/0007220.
\bibitem{Fierz:1939ix} 
M.~Fierz and W.~Pauli, 
Proc.\ Roy.\ Soc.\ Lond.\  {\bf A173}, 211 (1939). 
\bibitem{Wald1} C. Cutler and R. Wald, 
Class. Quant. Grav. {\bf 4} (1987) 1267. 
\bibitem{Wald2} R. Wald,  
Class. Quant. Grav. {\bf 4} (1987) 1279. 
\bibitem{ovrut} A. Hindawi, B. Ovrut, and D. Waldram, 
Phys. Rev. {\bf D53}  
(1996) 5583. 
\bibitem{bh} G.~Barnich and M.~Henneaux, 
Phys.\ Lett.\  {\bf B311}, 123 (1993) 
[hep-th/9304057];  
M.~Henneaux, 
hep-th/9712226. 
\bibitem{bv1} I.~A.~Batalin and G.~A.~Vilkovisky,
Phys.\ Lett.\  {\bf B102}, 27 (1981).
\bibitem{locality} M.~Henneaux,
Commun.\ Math.\ Phys.\  {\bf 140}, 1 (1991).
\bibitem{BBH1} G.~Barnich, F.~Brandt and M.~Henneaux,
Commun.\ Math.\ Phys.\  {\bf 174}, 57 (1995)
[hep-th/9405109].
\bibitem{BBH2} G.~Barnich, F.~Brandt and M.~Henneaux,
Commun.\ Math.\ Phys.\  {\bf 174}, 93 (1995)
[hep-th/9405194]; see also 
G.~Barnich and M.~Henneaux,
Phys.\ Rev.\ Lett.\  {\bf 72}, 1588 (1994)
[hep-th/9312206].
\bibitem{Guptaetal} S. N. Gupta,
Phys. Rev. {\bf 96} (1954) 1683;
R. H. Kraichnan,
Phys. Rev. {\bf 98} (1955) 1118;
R.~P.~Feynman, F.~B.~Morinigo, W.~G.~Wagner and B.~Hatfield,
``Feynman lectures on gravitation,''
{\it  Reading, USA: Addison-Wesley (1995)};
V. I. Ogievetsky and I. V. Polubarinov,
Ann. Phys. {\bf 35} (1965) 167;
S.~Deser,
Gen.\ Rel.\ Grav.\  {\bf 1}, 9 (1970);
F.~A.~Berends, G.~J.~Burgers and H.~Van Dam,
Z.\ Phys.\  {\bf C24}, 247 (1984);
R.~M.~Wald,
Phys.\ Rev.\  {\bf D33}, 3613 (1986).
\bibitem{ArDe} C.~Aragone and S.~Deser,
Nuovo Cim.\  {\bf B57}, 33 (1980).
\bibitem{Reuter} M. Reuter,
Phys. Rev. Lett. {\bf B205}
(1988) 511.
\end{thebibliography}
\end{document}